\documentclass[preprints,article,accept,moreauthors,pdftex]{Definitions/mdpi}
\firstpage{1} 
\pubvolume{xx}
\issuenum{1}
\articlenumber{5}
\pubyear{2019}
\copyrightyear{2019}
\history{Received: date; Accepted: date; Published: date}
\usepackage{rotating} % <-- HERE

\Title{Intracranial Hemorrhage Segmentation Using Deep Convolutional Model}

\Author{Murtadha D. Hssayeni$^{1,2}$, M.S., Muayad S. Croock$^{2}$, Ph.D., Aymen Al-Ani$^{2}$, Ph.D.,  \\Hassan Falah Al-khafaji$^{3}$, M.D., Zakaria A. Yahya$^{3}$, M.D. and Behnaz Ghoraani$^{1,*}$, Ph.D.}

\AuthorNames{Murtadha D. Hssayeni, Aymen Al-Ani, Hassan Falah Al-khafaji, Muayad S. Croock, Zakaria A. Yahya and Behnaz Ghoraani}

\address{%
$^{1}$ \quad The Department of Computer and Electrical Engineering and Computer Science, Florida Atlantic University, Boca Raton, FL 33431. \\
$^{2}$ \quad Computer Engineering Department, University of Technology, Baghdad, Iraq. \\
$^{3}$ \quad Babylon Health Directorate, Babil, Iraq. }

\corres{Correspondence: bghoraani@fau.edu.}

\abstract{ Traumatic brain injuries  could cause  intracranial hemorrhage (ICH). ICH could lead to disability or death if it is not accurately diagnosed and treated in a time-sensitive procedure.  The current clinical protocol to diagnose ICH is examining  Computerized Tomography (CT) scans by radiologists to detect ICH and localize its regions. However, this process  relies heavily  on the availability of an experienced radiologist. In this paper, we designed a study protocol to collect a dataset of 82 CT scans of subjects with traumatic brain injury. Later, the ICH regions were manually delineated in each slice by a consensus decision of  two radiologists. Recently, fully convolutional networks (FCN) have shown to be successful in medical image segmentation. We developed a deep FCN, called U-Net, to segment the ICH regions from the CT scans in a fully automated manner.  The method achieved a Dice coefficient of 0.31 for the ICH segmentation based on 5-fold cross-validation. The dataset is publicly available online at PhysioNet repository for future analysis and comparison. }   


\dataset{\href{https://physionet.org/content/ct-ich/1.2.0/}{https://physionet.org/content/ct-ich/1.2.0/}, doi:10.13026/w8q8-ky94}

\datasetlicense{Creative Commons Attribution 4.0 International Public License}

\begin{document}
%%%%%%%%%%%%%%%%%%%%%%%%%%%%%%%%%%%%%%%%%%

%%%%%%%%%%%%%%%%%%%%%%%%%%%%%%%%%%%%%%%%%%

    \section{Introduction}
        Traumatic brain injury (TBI) is a major cause of death and disability in the United States. It contributed to about 30\% of all injury deaths in 2013 \cite{taylor2017traumatic}. After accidents  with  TBI, extra-axial intracranial lesions, such as intracranial hemorrhage (ICH),  may occur. ICH is  a critical medical lesion that results in a high rate of mortality \cite{van2010incidence}. It is considered to be clinically dangerous because of its high risk for turning into a secondary brain insult that may lead to paralysis and even death if it is not treated in a time-sensitive procedure.  Depending on its location in the brain, ICH is divided into five sub-types:  Intraventricular (IVH), Intraparenchymal (IPH), Subarachnoid (SAH), Epidural (EDH) and Subdural (SDH).  In addition, the ICH that occurs within the brain tissue is called Intracerabral hemorrhage. 
        
        The Computerized Tomography (CT) scan is commonly used in the emergency evaluation of subjects with TBI  for ICH  \cite{currie2016imaging}. The availability of the CT scan and its rapid acquisition time makes it a preferred diagnostic tool over Magnetic Resonance Imaging for the initial assessment of ICH. CT scans generate a sequence of images using X-ray beams where brain tissues are captured with different intensities depending on the amount of X-ray absorbency (Hounsfield units (HU)) of the tissue. CT scans are displayed using a windowing method.  This method transforms the  HU numbers into grayscale values ([0, 255])  according to the window  level and width parameters.  By selecting different window parameters,  different features of the brain tissues are displayed in the grayscale image (e.g., brain window, stroke window, and bone window) \cite{xue2012window}. In the CT scan images based on the brain window,  the ICH regions  appear as hyperdense regions with a relatively undefined structure. These CT images are examined by an expert radiologist to determine whether ICH has occurred and if so, detect its type and region. However, this diagnosis process relies on the availability of a subspecialty-trained neuroradiologist, and as a result, could be time inefficient and even inaccurate, especially in remote areas where specialized care is scarce.

        Recent advances in convolutional neural networks (CNN) have demonstrated that the method has excellent performance in automating multiple image classification and segmentation tasks \cite{litjens2017survey}. Hence, we hypothesized that deep learning algorithms have the potential to automate the diagnosis procedure for segmenting the ICH regions. We developed a fully convolutional network (FCN) to segment the ICH regions in each CT slice. Such a method could help reducing the time and error in the ICH diagnosis significantly  where expert radiologists are not readily available . An automated ICH screen tool can be used to assist junior radiology trainees in detecting ICH and its sub-types, or when experts are not immediately available in the emergency rooms, especially in developing countries or remote areas.
        
        Furthermore, there is only one publicly available dataset called CQ500 for the detection of ICH sub-types \cite{chilamkurthy2018deep} that consists of 491 head CT scans.  There is no publicly available  dataset for the ICH segmentation. Hence, there is a need for a benchmark dataset  that could help to extend the work in  ICH segmentation. Therefore, the other focus of this work was collecting head CT scans with ICH segmentation and making it publicly available. We also performed a comprehensive literature review in the area of ICH detection and segmentation. 
        
    %%%%%%%%%%%%%%%%%%%%%%%%%%%%%%%%%%%%%%%%%%%%%%%%%%%%%%%%%%%%%%%%%%%%%%%%%%%%%%%%
    \section{Related work} \label{relatedWork}
\begin{figure}[hpb]
    		\centering
    		\includegraphics[width=0.7\linewidth]{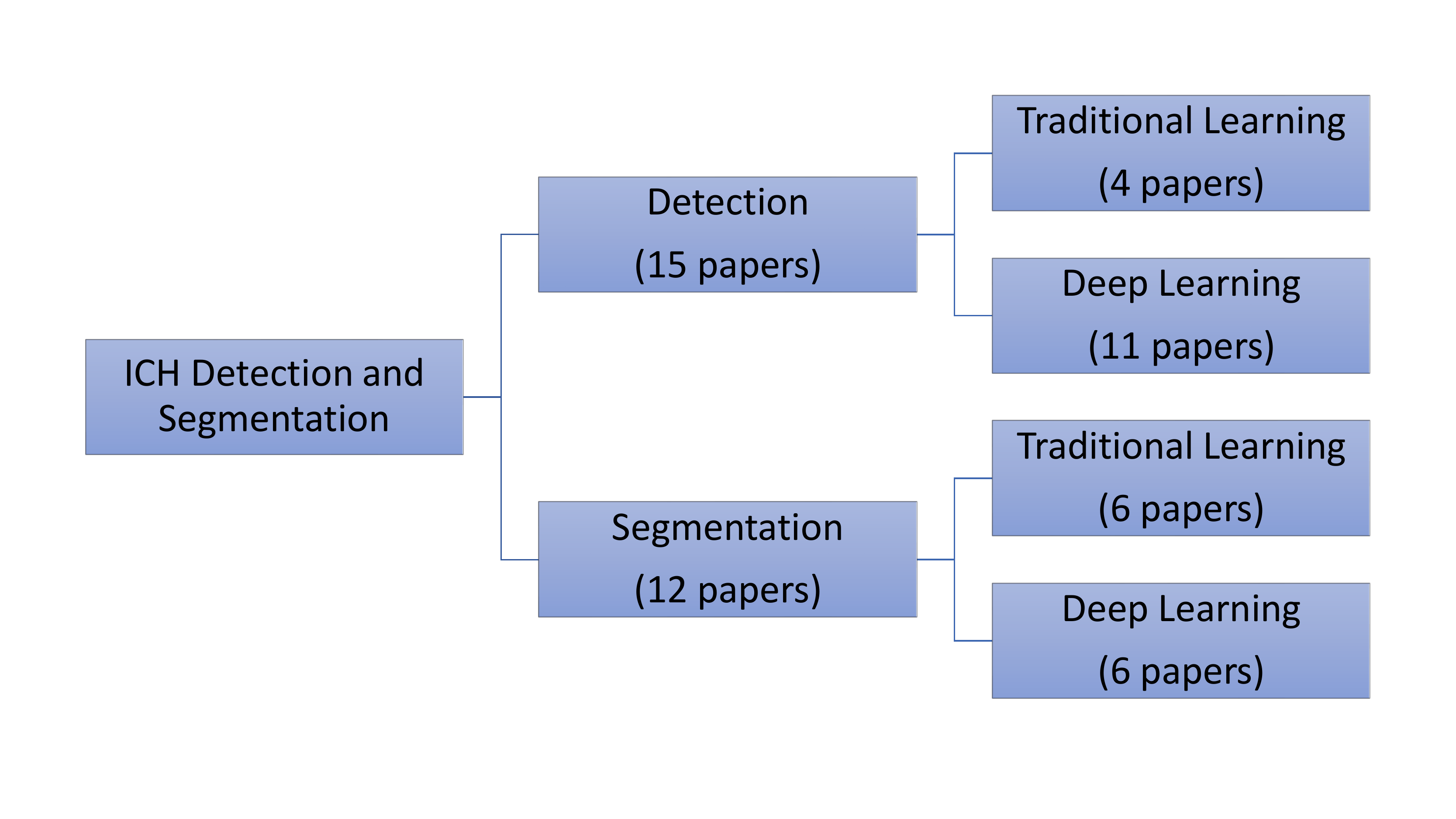}
    		\caption{The distribution of the reviewed papers for ICH detection and segmentation.}
    		\label{Fig:review}
    	\end{figure}

         Much interesting work has been performed for the automated ICH diagnosis. The majority of this work has focused either on a two-class detection problem where the method detects the presence of an ICH \cite{chan2007computer, yuh2008computer,li2010automatic, li2012automatic, shahangian2016automatic,prevedello2017automated, grewal2018radnet,jnawali2018deep, chang2018hybrid, arbabshirani2018advanced, chilamkurthy2018deep, arbabshirani2018advanced,lee2019explainable, ye2019precise, cho2019improving} or as a multi-class classification problem, where the goal is to detect the ICH sub-types \cite{yuh2008computer, shahangian2016automatic, chang2018hybrid, chilamkurthy2018deep, lee2019explainable, ye2019precise, cho2019improving}. Some researchers have extended the scope and performed the ICH segmentation to identify the region of ICH \cite{chan2007computer, prakash2012segmentation, bhadauria2014intracranial, shahangian2016automatic, muschelli2017pitchperfect, kuo_cost_sensitive_hemorrhage,chang2018hybrid,nag2018computer, lee2019explainable, kuang2019segmenting, cho2019improving, gautam2019automatic}. Most researchers validated their algorithms using small datasets \cite{chan2007computer, yuh2008computer,li2010automatic, li2012automatic,prakash2012segmentation, bhadauria2014intracranial,shahangian2016automatic, muschelli2017pitchperfect, prevedello2017automated, grewal2018radnet, nag2018computer,lee2019explainable, kuang2019segmenting, gautam2019automatic}, while a few used large datasets for testing and validating \cite{jnawali2018deep, arbabshirani2018advanced, kuo_cost_sensitive_hemorrhage, chilamkurthy2018deep, chang2018hybrid, ye2019precise, cho2019improving}. We provide a comprehensive review of the published papers for the ICH detection and segmentation (Figure \ref{Fig:review}) in this section.
         
        \subsection{Intracranial Hemorrhage Detection}
    	    Several traditional and deep learning approaches were developed in the literature. Regarding the traditional machine learning methods, Yuh and colleagues developed a threshold-based algorithm to detect ICH. Later, the method detected the ICH sub-types based on its location, shape, and volume  \cite{yuh2008computer}. The authors optimized the value of the threshold using the retrospective samples of 33 CT scans and evaluated their model on 210 CT scans of subjects with suspected TBI. Their algorithm achieved 98\% sensitivity and 59\% specificity for  the  ICH detection and an intermediate accuracy in detecting the ICH sub-types. In another work, Li and colleagues proposed two methods to segment the SAH space and then used the segmented regions to detect the SAH hemorrhage \cite{li2010automatic, li2012automatic}. One method used elastic registration with  the  SAH space atlas, whereas the other method extracted distance transform features and trained a Bayesian decision method to perform the delineation. After  the  SAH space segmentation, mean gray value, variance, entropy, and energy were extracted and used to train a support vector machine classifier for  the  SAH hemorrhage detection. They used 60 CT scans (30 with SAH hemorrhage) to train the algorithm and tested the model on 69 CT scans (30 with SAH hemorrhage). The best performance was reported using the Bayesian decision method with 100\% testing sensitivity, 92\% specificity, and 91\% accuracy \cite{li2012automatic}.

    	    Regarding the deep learning approaches, all  the methods were based on CNN and its variants  except for the approaches in Refs.  \cite{kuo_cost_sensitive_hemorrhage, kuang2019segmenting, cho2019improving} , which were based on  a FCN model. In these approaches, the spatial dependency between  the  adjacent slices was considered using a second model such as random forest \cite{chilamkurthy2018deep}  or  RNN \cite{grewal2018radnet,ye2019precise}. Some authors also modified CNN to process  some  part or the entire CT scan \cite{jnawali2018deep,chang2018hybrid} or used an interpolation layer \cite{lee2019explainable}. Other approaches were 1-stage , meaning that they  did not consider the spatial dependency between the slices \cite{prevedello2017automated, kuo_cost_sensitive_hemorrhage,cho2019improving}. Prevedello and colleagues proposed two algorithms based on CNN \cite{prevedello2017automated}. One  of their algorithms  was focused on detecting ICH, mass effect,  and  hydrocephalus at  the  CT scans while their other  algorithm  was developed to detect  the  suspected acute infarcts. A total of 246 CT scans were used for training and validation (100 hydrocephalus, 22 suspected acute infarct, and 124 noncritical findings), and a total of 100 CT scans were used for testing (50 hydrocephalus, 15 SAI, and 35 noncritical findings). The testing predictions were validated with the final radiology report or with the neuroradiologist' review for  the  equivocal findings. The hydrocephalus detection algorithm yielded 90\% sensitivity, 85\% specificity, and  the  area under the curve (AUC) of 0.91. The suspected acute infarct detection algorithm resulted in  a  lower specificity and AUC of 0.81.
    
    		Chilamkurthy and colleagues proposed four algorithms to detect the sub-types of ICH, calvarial fractures, midline shift, and mass effect \cite{chilamkurthy2018deep}. They trained and validated the algorithms on  a  large dataset with 290k and 21k CT scans, respectively. Two datasets were used for testing. A part of the testing,  a  dataset with 491 scans was made public (called CQ500). Clinical radiology reports were used as the gold standard to label the training and validation  CT  scans. These reports were used to label each scan utilizing a natural language processing algorithm. The testing scans were then annotated by the majority vote of the ICH sub-types reported by three expert radiologists. Different deep models were developed for each of the four categories. ResNet18 was trained with five parallel fully connected layers as the output layers. The results of these output layers for each slice were fed to  a  random forest algorithm to predict  the  scan-level confidence for the presence of  an ICH . They reported an average AUC of 0.93 for the ICH sub-type detection on both datasets. Considering the high sensitivity operating point, the average sensitivity was 92\%, which was similar to that of the radiologists. However, the average specificity was 70\%, which was significantly lower than the golden standard. Also, it varied for different ICH sub-types. The lowest specificity of 68\% was for the SDH detection.
        		
    		Two approaches  based on CNN with RNN  were proposed to detect ICH \cite{grewal2018radnet, ye2019precise}. Grewal \textit{et al.} \cite{grewal2018radnet} proposed a 40-layer CNN, called DenseNet, with Bidirectional long short-term memory (LSTM) layer for  the  ICH detection.  They also introduced  three auxiliary tasks after each Dense Convolutional block to compute  the  binary segmentation of  the ICH  regions. Each of these tasks consisted of one convolutional layer followed by  a  deconvolution layer  in order to  upsample the feature maps to the original image size.  The  LSTM layer was added to incorporate the inter-slice dependencies of the CT scans of each  subject . They considered 185 CT scans for training, 67 for validation, and 77 for testing. The training data was augmented by rotation and horizontal flipping to balance the number of scans for each of the two classes. The network detection of the test data was evaluated against the annotation of three expert radiologists for each CT slice. They reported 81\% accuracy, 88\% recall (sensitivity), 81\% precision, and 84\% F1 score. The model F1 score was higher than two of the three radiologists. Also, adding attention layers provided a significant increase in  the  model sensitivity. In \cite{ye2019precise}, the authors presented a 3D joint convolutional and recurrent neural network (CNN-RNN) to detect and classify \cite{ye2019precise} ICH regions. The overall architecture of this model was similar to the model proposed by Grewal \textit{et al.} \cite{grewal2018radnet}. VGG-16 was used as the CNN model, and bidirectional Gated Recurrent Unit (GRU) was used as the RNN model. RNN layer had the same functionality of the slice interpolation technique proposed by \cite{lee2019explainable}, but it was more flexible  with respect to  the number of adjacent slices included in the classification. The algorithm was trained and validated on 2,537 CT scans and tested on 299 CT scans. They reported  a  precise slice-level ICH detection with 99\% for both sensitivity and specificity and  an  AUC of 1. However, for classification of the ICH sub-types, they reported  a  lower performance with 80\% average sensitivity, 93.2\% average specificity, and  an  AUC of 0.93. The lowest sensitivity was reported for SAH and EDH, which was 69\% for both sub-types.

            In three approaches, the CNN model was modified to process a number of CT slices at once \cite{jnawali2018deep, chang2018hybrid, arbabshirani2018advanced}. Jnawalia and colleagues \cite{jnawali2018deep} proposed an ensemble of three different CNN models to perform  the  ICH detection. The CNN models were based on the architectures of AlexNet and GoogleNet that were  extended to a 3D model by taking all the slices for each CT scan. They also have  a lower number of parameters by reducing the number of layers and filter specifications. They trained, validated, and tested their model on a large dataset with 40k CT scans. About 34k CT scans were used for training (26K normal scans).  However , the method that was used to label the CT scans was not reported. The positive slices were oversampled and augmented to make a balanced training dataset. About 2k and 4k scans were used for validation and testing, respectively. The AUC of the ensemble of the CNN models was 87\% with the precision of 80\%, recall of 77\%, and F1-score of 78\%.
            Chang and colleagues also developed a deep learning algorithm to detect ICH and its sub-types (except for IVH) with an ability to segment the ICH  regions  and quantify the ICH volume \cite{chang2018hybrid}. Their deep model is based on  a  region-of-interest CNN that estimates regions that contain  an  ICH for each five CT slices and then generates a segmentation mask for  the  positive cases of ICH. The authors trained their algorithm on a dataset with 10k CT scans and tested it on a prospective dataset of 862 CT scans. The reported 95\% sensitivity, 97\% specificity, and  an  AUC of 0.97 for the classification of ICH sub-types and an average Dice score of 0.85 for the ICH segmentation. The lowest detection sensitivity of 90\% and Dice score of 0.77 were reported for SAH.
             In \cite{arbabshirani2018advanced}, an ensemble of four 3D CNN models with an input shape of $24\times256\times256$ was implemented and evaluated using 9,499 retrospective and 347 prospective CT scans. An AUC of 0.846 was achieved on the retrospective study, and an average sensitivity of 71.5\% and specificity of 83.5\% were obtained on both testing datasets.

    		Similar to the work of Jnawalia \textit{et al.} \cite{jnawali2018deep}, Lee and colleagues used transfer learning on an ensemble of four well-known CNN models to detect  the  ICH sub-types and bleeding points \cite{lee2019explainable}. The four models were VGG-16, ResNet-50, Inception-v3, and Inception-ResNet-v2. the spatial dependency between  the  adjacent slices  was taken into consideration by  introducing a slice interpolation technique. This ensemble model was trained and validated  using  a dataset with 904 CT scans and tested  using  a retrospective dataset with 200 CT scans and a prospective dataset with 237 scans. On average, the ICH detection algorithm resulted a testing AUC  of 0.98  with 95\%  sensitivity and specificity. However,  the  algorithm resulted in  a  significantly lower sensitivity for the classification of the ICH sub-types with 78.3\%  sensitivity and  92.9\% specificity. The lowest sensitivity of 58.3\% was reported for the EDH  slices  in the retrospective test set and 68.8\% for the IPH  slices  in the prospective test set. The overall localization accuracy of the attention maps was 78.1\% between the model segmentation and  the  radiologists' maps of bleeding points.

        \subsection{Intracranial Hemorrhage Segmentation}
    		It is essential to localize and find the ICH volume to decide on the appropriate medical and surgical intervention \cite{chi2014relationship}. Several methods were proposed to automate the process of the ICH segmentation \cite{chan2007computer, prakash2012segmentation, bhadauria2014intracranial, shahangian2016automatic, muschelli2017pitchperfect, kuo_cost_sensitive_hemorrhage,chang2018hybrid,nag2018computer, lee2019explainable, kuang2019segmenting, cho2019improving, gautam2019automatic}. 
    		 Similar to the ICH detection, the ICH delineation approaches can be divided into traditional \cite{chan2007computer, prakash2012segmentation, bhadauria2014intracranial,shahangian2016automatic, muschelli2017pitchperfect, gautam2019automatic} and deep learning methods \cite{chang2018hybrid,lee2019explainable,kuo_cost_sensitive_hemorrhage, nag2018computer, kuang2019segmenting, cho2019improving}. 
    		
    		The traditional methods usually require preprocessing of the CT scans to remove the skull and noise. They also require to register the segmented brains and extract some complicated engineered features. Many of these methods are based on unsupervised clustering to segment the ICH regions \cite{prakash2012segmentation, bhadauria2014intracranial, shahangian2016automatic, gautam2019automatic}. The methods in Ref. \cite{prakash2012segmentation} and \cite{shahangian2016automatic} both use the Distance Regularized Level Set Evolution (DRLSE) method to fit active contours on ICH regions.
            Prakash and colleagues modified DRLSE for the segmentation of the IVH and IPH regions after preprocessing the CT scans for the skull removal and noise filtering \cite{prakash2012segmentation}. Validating the method on 50 test CT scans resulted in an average Dice coefficient of 0.88, 79.6\% sensitivity, and 99.9\% specificity.
            Shahangian and colleagues used DRLSE for the segmentation of the EDH, IPH and SDH regions and also proposed a supervised method based on support vector machine for the classification of the ICH slices \cite{shahangian2016automatic}. The first step in their method was segmenting the brain by removing the skull and brain ventricles. Next, they performed the ICH segmentation based on DRLSE. Then, they extracted the shape and texture features of the ICH regions, and finally, they performed the ICH detection. This method resulted in an average Dice coefficient of 58.5, 82.5\% sensitivity, and 90.5\% specificity on 627 CT slices.
            The other traditional unsupervised studies \cite{bhadauria2014intracranial} \cite{gautam2019automatic} used a fuzzy c-means clustering approach for the ICH segmentation. The authors in Ref.  \cite{bhadauria2014intracranial} proposed a method based on a spatial fuzzy c-means clustering and region-based active contour model. A retrospective set of 20 CT scans with  an  ICH was used. The authors reported 79\% sensitivity, 99\% specificity, and an average Jaccard index of 0.78. Similarly,  Gautam and colleagues proposed a method based on the white matter fuzzy c-means clustering followed by a wavelet-based thresholding technique \cite{gautam2019automatic}. They evaluated their method on 20 CT scans with  an  ICH and reported a Dice coefficient of 0.82. 

            Unlike the unsupervised methods, 
        the traditional supervised approaches \cite{chan2007computer, muschelli2017pitchperfect} use labeled slices to train the classifiers.  The authors in  \cite{chan2007computer} proposed a semi-automatic ICH segmentation method where the brain in each CT slice was first segmented and aligned. Then,  the  candidate ICH regions  were  selected using top-hat transformation and extraction of  the  asymmetrical high intensity regions. Finally, the candidate regions were fed to a knowledge-based classifier for  the  ICH detection.  This method resulted in  100\% slice-level sensitivity, 84.1\% slice-level specificity, and 82.6\% lesion-level sensitivity. In another work, Muschelli and colleagues \cite{muschelli2017pitchperfect} proposed a fully-automatic method. They compared multiple traditional supervised methods  for  the segmentation of intracerebral hemorrhage \cite{muschelli2017pitchperfect}.  For this purpose , the brains were first extracted from the CT scans and registered using a CT brain-extracted template. Next, multiple features were extracted from each scan.  The features consisted of  threshold-based information of the CT voxel intensity, local moment information, such as mean and std, within-plane standard scores, initial segmentation using an unsupervised model, contralateral difference images, distance to the brain center, and the standardized-to-template intensity that contrast a given CT scan with an averaged CT scans from healthly subjects. The classification  models considered in this study were  logistic regression, generalized additive model, and random forest.  These models  were trained on 10 CT scans and tested on 102 CT scans. Random forest resulted in the highest Dice coefficient of 0.899.

    		The deep learning approaches for the ICH segmentation were either based on CNN \cite{chang2018hybrid,lee2019explainable, nag2018computer} or the FCN design \cite{kuo_cost_sensitive_hemorrhage, kuang2019segmenting, cho2019improving}. In the previous section, two methods for  the  ICH segmentation based on CNN were reviewed \cite{chang2018hybrid,lee2019explainable}. Another work was developed by Nag and colleagues where the authors first selected the CT slices with  an  ICH using a trained autoencoder and then segmented the ICH areas using the active contour Chan-Vese model \cite{nag2018computer}. A dataset with 48 CT scans was used to evaluate the method. The autoencoder was trained on half of the data  and all the data  was used to test the algorithm. This work reported a sensitivity of 71\%, positive predictive value of 73\%, and Jaccard index of 0.55.
    		
    		FCN provides an ability to  predict  the presence of ICH at the pixel level.  This ability of FCN  can also be used for the ICH segmentation.  Several architectures of FCN were used for  the  ICH segmentation as follows: dilated residual net (DRN) \cite{kuo_cost_sensitive_hemorrhage}, modified VGG16 \cite{cho2019improving}, and U-Net \cite{kuang2019segmenting}.  
    		The authors in Ref. \cite{kuo_cost_sensitive_hemorrhage} proposed a cost-sensitive active learning system. The system consisted of  the  ensemble of  a  patch fully CNN (PatchFCN). After the PatchFCN,  the  uncertainty score was calculated for each patch, and the sum of these patches' scores was maximized under the estimated labeling time constraint. The authors used 934 CT scans for training and validating purposes and 313  retrospective scans and 100 prospective scans for testing purposes. They reported 92.85\% average precision for  the  ICH detection at scan level using both test sets and 77.9\%  average precision for  the  segmentation. The application of the cost-sensitive active learning technique improved the model performance on the prospective test set by annotating the new CT scans and increasing the size of the training data/scans. 
    		 In \cite{cho2019improving}, the CNN cascade model was used for the ICH detection and the dual FCN models was used for the ICH segmentation. The CNN cascade model was based on the GoogLeNet network, and the dual FCN model was based on a pre-trained VGG16 network that was modified and fine-tuned on the brain and stroke window settings. The methods were evaluated using 5-fold cross-validation of about 6k CT scans. The authors reported a sensitivity and specificity of about 98\% for the ICH binary classification and an accuracy ranging from 70\% to 90\% for the ICH sub-type detection. The lowest accuracy was reported for the EDH detection. For the ICH segmentation, they reported 80.19\% precision and 82.15\% recall.
    		Kuang and colleagues proposed a semi-automatic method to segment the regions of intracerebral hemorrhage in addition to the ischemic infarct segmentation \cite{kuang2019segmenting}. The method consisted of U-Net models for the ICH and infarct segmentation that was fed beside a user initialization of the ICH and infarct regions to a multi-region contour evolution. A set of hand-crafted features based on the bilateral density difference between the symmetric brain regions in the CT scan was introduced into the U-Net. Also, the authors weighted the U-Net cross-entropy loss by the Euclidean distance between a given pixel and the boundaries of the true masks. The proposed semi-automatic method with the weighted loss outperformed the traditional U-net where it achieved a Dice similarity coefficient of 0.72.  
    		
    		%The studies of Muschelli and Kuang was focusing on the segmentation of only intracerebral hemorrhage. Cho and colleagues proposed a deep learning method for the detection and segmentation of all ICH types.

    	 Table \ref{Table:Review1} and \ref{Table:Review2} summarize  the methods for  the  ICH detection and segmentation.  As expected, a  high testing sensitivity and specificity was reported on large datasets, and the  performance of the ICH  detection algorithms was equivalent to  the results from the  senior expert radiologists \cite{chilamkurthy2018deep,chang2018hybrid,lee2019explainable,ye2019precise}. However, the sensitivity of the detection of some ICH sub-types was equivalent to the  the results from the  junior radiology trainees \cite{ye2019precise}. SAH and EDH were the most difficult ICH sub-types to be classified by all the machine learning models \cite{chang2018hybrid,lee2019explainable, ye2019precise, kuang2019segmenting}.  It is interesting to note that  SAH is also reported to be the most miss-classified sub-type by radiology residents \cite{strub2007overnight}. For  the  ICH segmentation, the machine learning methods achieved  a relatively high performance  \cite{prakash2012segmentation, bhadauria2014intracranial, muschelli2017pitchperfect, kuo_cost_sensitive_hemorrhage,chang2018hybrid, cho2019improving, kuang2019segmenting,  gautam2019automatic}. However, there is still a need for a method that can precisely delineate the regions of all ICH sub-types. To address this need, we first collected a dataset of CT scans,  which is available online at \href{https://physionet.org/content/ct-ich/1.2.0/}{https://physionet.org/content/ct-ich/1.2.0/}. Then, we implemented a fully convolutional network,  known as U-Net,  for  the  ICH segmentation.

        \begin{sidewaystable}[htbp!]
        \centering
        \caption{Review of the methods proposed for  the  ICH detection and segmentation. Some papers used retrospective and prospective sets to test their models (i.e., retrospective + prospective), so the reported results are the average of both sets.}
        \label{Table:Review1}
        \resizebox{0.98\linewidth}{!}{\begin{tabular}{|l|l|l|l|l|l|l|l|l|l|}
\hline
\multirow{3}{*}{\textbf{References}} & \multicolumn{4}{l|}{\textbf{Dataset (\# of CT scans)}}                                                                                                     & \multirow{3}{*}{\textbf{\begin{tabular}[c]{@{}l@{}}ICH Detection\\ Method\end{tabular}}}                                                   & \multicolumn{2}{l|}{\multirow{2}{*}{\textbf{Results}}}                                                                                                                                                                                             & \multirow{3}{*}{\textbf{\begin{tabular}[c]{@{}l@{}}ICH Segmentation\\ Method\end{tabular}}}         & \multirow{2}{*}{\textbf{Results}}                                                                                            \\ \cline{2-5}
                                     & \multicolumn{2}{l|}{\textbf{Training}}   & \multicolumn{2}{l|}{\textbf{Testing}}                                                                           &                                                                                                                                            & \multicolumn{2}{l|}{}                                                                                                                                                                                                                              &                                                                                                     &                                                                                                                              \\ \cline{2-5} \cline{7-8} \cline{10-10} 
                                     & \textbf{With ICH} & \textbf{Without ICH} & \textbf{With ICH}                                     & \textbf{Without ICH}                                    &                                                                                                                                            & \textbf{ICH}                                                                                                                       & \textbf{ICH Sub-types}                                                                                        &                                                                                                     & \textbf{ICH Segmentation}                                                                                                    \\ \hline
\textit{Yuh et. al. \cite{yuh2008computer}}                 & 27                & 5                    & 52                                                    & 158                                                     & Threshold-based                                                                                                                            & \multicolumn{2}{l|}{\begin{tabular}[c]{@{}l@{}}98\% sensitivity \\ 59\% specificity\end{tabular}}                                                                                                                                                  & Threshold-based                                                                                     & -                                                                                                                            \\ \hline
\textit{Li et. al. \cite{li2010automatic, li2012automatic}}                  & 30                & 30                   & 30                                                    & 39                                                      & \begin{tabular}[c]{@{}l@{}}Support Vector \\ Machine\end{tabular}                                                                          & -                                                                                                                                  & \begin{tabular}[c]{@{}l@{}}100\% sensitivity \\ 92\% specificity\\ (SAS detection)\end{tabular}               & \begin{tabular}[c]{@{}l@{}}Distance transform \\ features and a Bayesian \\ classifier\end{tabular} & -                                                                                                                            \\ \hline
\textit{Prevedello et. al. \cite{prevedello2017automated}}          & 100               & 124                  & 50                                                    & 35                                                      & \begin{tabular}[c]{@{}l@{}}Convolutional Neural\\ Networks\end{tabular}                                                                    & \begin{tabular}[c]{@{}l@{}}90\% sensitivity\\ 85\% specificity\\ AUC of 0.91\end{tabular}                                          & -                                                                                                             & -                                                                                                   & -                                                                                                                            \\ \hline
\textit{Grewal et. al.\cite{grewal2018radnet}}              & \multicolumn{2}{l|}{252}                 & \multicolumn{2}{l|}{77}                                                                                         & \begin{tabular}[c]{@{}l@{}}Convolutional Neural\\ Networks (DenseNet) \\ + RNN\end{tabular}                                                & \begin{tabular}[c]{@{}l@{}}88\% sensitivity\\ 81\% precision\\ 81\% accuracy\end{tabular}                                          & -                                                                                                             & \begin{tabular}[c]{@{}l@{}}Auxiliary tasks to \\ DenseNet\end{tabular}                              & -                                                                                                                            \\ \hline
\textit{Jnawali et. al. \cite{jnawali2018deep}}             & 8,465             & 26,383               & 1,891                                                 & 3,618                                                   & \begin{tabular}[c]{@{}l@{}}Convolutional Neural\\ Networks (ensemble)\end{tabular}                                                         & \begin{tabular}[c]{@{}l@{}}77\% sensitivity\\  80\% precision\\ AUC of 0.87\end{tabular}                                           & -                                                                                                             & -                                                                                                   & -                                                                                                                            \\ \hline
\textit{Chilamkurthy et. al. \cite{chilamkurthy2018deep}}        & \multicolumn{2}{l|}{290,055}             & \begin{tabular}[c]{@{}l@{}}2,494\\ +205\end{tabular}& \begin{tabular}[c]{@{}l@{}}18,601\\+286\end{tabular}& \begin{tabular}[c]{@{}l@{}}Convolutional Neural\\ Networks (ResNet18)\\ and Random Forest\end{tabular}                                     & \multicolumn{2}{l|}{\begin{tabular}[c]{@{}l@{}}92\% sensitivity \\ 70\% specificity\\ Average AUC of 0.93 (All types)\end{tabular}}                                                                                                                & -                                                                                                   & -                                                                                                                            \\ \hline
\textit{Arbabshirani at. al. \cite{arbabshirani2018advanced}}        & 9,938             & 27,146               & \multicolumn{2}{l|}{9,499+347}                                                                                  & 3D CNN                                                                                                                                     & \begin{tabular}[c]{@{}l@{}}AUC of 0.846\\ 71.5\% sensitivity\\ 83.5\% specificity\end{tabular}                                     & -                                                                                                             & -                                                                                                   & -                                                                                                                            \\ \hline
\textit{Ye et. al. \cite{ye2019precise}}                  & 1,642             & 895 & 194 & 105                                                     & 3D joint CNN-RNN                                                                                                                           & \begin{tabular}[c]{@{}l@{}}98\% sensitivity\\ 99\% specificity\\ AUC of 1\end{tabular}                                             & \begin{tabular}[c]{@{}l@{}}80 \% sensitivity \\ 93.2 \% specificity\\ AUC of  0.93\\ (All types)\end{tabular} & \begin{tabular}[c]{@{}l@{}}Attention maps\\ of CNN using\\ Grad-CAM method\end{tabular}             & -                                                                                                                            \\ \hline
\textit{Chan \cite{chan2007computer}}                                 & 40                & 124                  & 22                                                    & 0                                                       & Knowledge-based classifier                                                                                                                 & \begin{tabular}[c]{@{}l@{}}100\% sensitivity\\ 84.1\% specificity\end{tabular}                                                     & -                                                                                                             & \begin{tabular}[c]{@{}l@{}}Knowledge-based \\ classifier\end{tabular}                               & 82.6\% sensitivity                                                                                                           \\ \hline
\textit{Shahangian at. el. \cite{shahangian2016automatic}}          & \multicolumn{2}{l|}{627 slices}          & 0                                                     & 0                                                       & Support Vector Machine                                                                                                                     & 92.46\% accuracy                                                                                                                   & 94.13\% accuracy                                                                                              & \begin{tabular}[c]{@{}l@{}}Distance regularized\\ level set evolution\end{tabular}                  & \begin{tabular}[c]{@{}l@{}}Dice coefficient of 58.5\\ 82.5\% sensitivity\\ 90.5\% specificity\end{tabular}                   \\ \hline
\textit{Chang et. al. \cite{chang2018hybrid}}               & 901               & 9,258                & 82                                                    & 780 & \begin{tabular}[c]{@{}l@{}}ROI-based Convolutional \\ Neural Networks\end{tabular}                                                         & \begin{tabular}[c]{@{}l@{}}95\% sensitivity \\ 97\% specificity\\ AUC of 0.97\\ (All types except\\ intraventricular)\end{tabular} & -                                                                                                             & \begin{tabular}[c]{@{}l@{}}ROI-based\\ Convolutional \\ Neural Networks\end{tabular}                & \begin{tabular}[c]{@{}l@{}}Average Dice\\ score of 0.85\end{tabular}                                                         \\ \hline
\textit{Lee et. al. \cite{lee2019explainable}}                 & 625               & 279                  & \begin{tabular}[c]{@{}l@{}}100\\ +\\ 107\end{tabular} & \begin{tabular}[c]{@{}l@{}}100\\ +\\ 130\end{tabular}   & \begin{tabular}[c]{@{}l@{}}Convolutional Neural\\ Networks (ensemble)\end{tabular}                                                         & \begin{tabular}[c]{@{}l@{}}95.2\% sensitivity \\ 94.9\% specificity\\ AUC of 0.975\end{tabular}                                    & \begin{tabular}[c]{@{}l@{}}78.3\% sensitivity \\ 92.9\% specificity\\ AUC of 95.9 \\ (All types)\end{tabular} & \begin{tabular}[c]{@{}l@{}}Attention maps\\ of CNN\end{tabular}                                     & \begin{tabular}[c]{@{}l@{}}78.1\% overlap between the \\ model and neuroradiologists \\ maps of bleeding points\end{tabular} \\ \hline
\textit{Kuo et. al. \cite{kuo_cost_sensitive_hemorrhage}}                 & \multicolumn{2}{l|}{934}                 & \multicolumn{2}{l|}{313+120}                                                                                    & \begin{tabular}[c]{@{}l@{}}Fully Convolutional \\ Neural Network (FCN)\end{tabular}                                                        & \begin{tabular}[c]{@{}l@{}}92.8\%  average \\ precision\end{tabular}                                                               & -                                                                                                             & \begin{tabular}[c]{@{}l@{}}Fully Convolutional \\ Neural Network (FCN)\end{tabular}                 & 77.9\%  average precision                                                                                                    \\ \hline
\textit{Cho et. al. \cite{cho2019improving}}                 & 2,647             & 3,055                & 0                                                     & 0                                                       & \begin{tabular}[c]{@{}l@{}}Cascade of convolutional\\ neural networks (CNN) \\ and dual fully convolutional \\ networks (FCN)\end{tabular} & \begin{tabular}[c]{@{}l@{}}97.91\% sensitivity\\ 98.76\% specificity\end{tabular}                                                  & \begin{tabular}[c]{@{}l@{}}Accuracy ranging\\ from 70\% to 90\%\end{tabular}                                  & \begin{tabular}[c]{@{}l@{}}Dual fully convolutional \\ networks (FCN)\end{tabular}                  & \begin{tabular}[c]{@{}l@{}}80.19\% precision\\ 82.15\% recall\end{tabular}                                                   \\ \hline
\end{tabular}}
        \end{sidewaystable}

            \begin{table}[htbp!]
        \centering
        \caption{Review of the methods proposed for  the  ICH segmentation only.}
        \label{Table:Review2}
        \resizebox{1\linewidth}{!}{\begin{tabular}{|l|l|l|l|l|l|l|}
\hline
\multirow{3}{*}{\textbf{References}} & \multicolumn{4}{l|}{\textbf{Dataset (\# of CT scans)}}                              & \multirow{3}{*}{\textbf{\begin{tabular}[c]{@{}l@{}}ICH Segmentation\\ Method\end{tabular}}}                                                                    & \multirow{2}{*}{\textbf{Results}}                                                                                                   \\ \cline{2-5}
                                     & \multicolumn{2}{l|}{\textbf{Training}}   & \multicolumn{2}{l|}{\textbf{Testing}}    &                                                                                                                                                                &                                                                                                                                     \\ \cline{2-5} \cline{7-7} 
                                     & \textbf{With ICH} & \textbf{Without ICH} & \textbf{With ICH} & \textbf{Without ICH} &                                                                                                                                                                & \textbf{ICH Segmentation}                                                                                                           \\ \hline
\textit{Bhadauria et. al. \cite{bhadauria2014intracranial}}           & 0                 & 0                    & 20                & 0                    & \begin{tabular}[c]{@{}l@{}}Fuzzy c-mean clustering \\ and region-based active \\ contour method\end{tabular}                                                   & \begin{tabular}[c]{@{}l@{}}79.4\% sensitivity \\ 99.4\% specificity\\ Jaccard index of 0.78\\ Dice coefficient of 0.87\end{tabular} \\ \hline
\textit{Nag et. al. \cite{nag2018computer}}                 & 24                & 0                    & 48                & 0                    & \begin{tabular}[c]{@{}l@{}}Autoencoder and active\\  contour Chan-Vese model\end{tabular}                                                                      & \begin{tabular}[c]{@{}l@{}}71\% sensitivity\\ 73\% positive predictive\\ Jaccard index of 0.55\end{tabular}                         \\ \hline
\textit{Muschelli et. al. \cite{muschelli2017pitchperfect}}           & 10                & 0                    & 102               & 0                    & \begin{tabular}[c]{@{}l@{}}Logistic regression, logistic \\ regression with LASSO, \\ Generalized additive model, \\ and random forest classifier\end{tabular} & \begin{tabular}[c]{@{}l@{}}Dice coefficient of 0.89\\ ICH volume correlation of 0.93\end{tabular}                                   \\ \hline
\textit{Kuang et. al. \cite{kuang2019segmenting}}               & 180               & 0                    & 30                & 0                    & \begin{tabular}[c]{@{}l@{}}U-Net and multi-region \\ contour evolution\end{tabular}                                                                            & Dice coefficient of 0.72                                                                                                            \\ \hline
\textit{Gautam et. al. \cite{gautam2019automatic}}              & 20                & 0                    & 0                 & 0                    & \begin{tabular}[c]{@{}l@{}}Fuzzy c-Mean clustering with\\ wavelet based thresholding \\ technique\end{tabular}                                                 & Dice coefficient of 0.82                                                                                                            \\ \hline
\textit{Prakash at. al. \cite{prakash2012segmentation}}             & 150               & 0                    & 50                & 0                    & \begin{tabular}[c]{@{}l@{}}Distance regularized\\ level set evolution\end{tabular}                                                                             & \begin{tabular}[c]{@{}l@{}}AUC of 0.88\\ 79.6\% sensitivity\\ 99.9\% specificity\end{tabular}                                       \\ \hline
\end{tabular}}
        \end{table}

        %\footnotetext{The paper has two test sets (retrospective + prospective) and the reported results are the average of both sets.}
        
    %%%%%%%%%%%%%%%%%%%%%%%%%%%%%%%%%%%%%%%%%%%%%%%%%%%%%%%%%%%%%%%%%%%%%%%%%%%%%%%%%%%%%%%%%%%%%%%%%%%%%%%%%%%%%%%%%%
    
    \section{Dataset} \label{dataset}
        A retrospective study was designed to collect head CT scans of subjects with TBI. The study was approved by the research and ethics board in the Iraqi ministry of health-Babil Office. The CT scans were collected between February and August 2018 from Al Hilla Teaching Hospital-Iraq. The CT scanner was Siemens/ SOMATOM Definition AS which had an isotropic resolution of 0.33 mm, 100 kV, and a slice thickness of 5mm. The information of each subject was anonymized. A total of 82 subjects (46 male) with an average age of 27.8$\pm$19.5 years were included in this study (refer to Table \ref{demo} for the subject demographics). Each CT scan includes about 30 slices. Two radiologists annotated the non-contrast CT scans and recorded the ICH sub-types if an  ICH was diagnosed. The two radiologists reviewed the non-contrast CT scans together and at the same time. Once they reached a consensus on the ICH diagnosis, which consisted of the presence of the ICH and its shape and location, the delineation of the ICH regions was performed to reduce the effort and time in the ICH segmentation process. The radiologists did not have access to the clinical history of the subjects. 
        
        During the data collection process, Syngo by Siemens Medical Solutions was first used to read the CT DICOM files and save two videos (AVI format), one using the brain window  (\color{blue}level=40\color{black}, width=120) and one using the bone window (level=700, width=3200). Second, a custom tool was implemented in Matlab and used to  perform the following tasks: 
        reading the AVI files, switching between the two window level settings, navigating between the slices, recording the radiologist annotations, delineating the ICH regions, and saving them as the binary 650x650 masks (JPG format). The gray-scale 650x650 images (JPG format) for each CT slice were also saved for both brain and bone windows \color{blue}(please refer to the supplement document for more details about the data collection process)\color{black}. 
        
        \begin{table}[ht]
        \centering
        \caption{Subject demographics.}
        \label{demo}
        \begin{tabular}{|l|l|l|l|}
        \hline
        \textbf{Total number of subjects}                                                                          & 82            & \textbf{Sex (Male, Female)}                                                                & 46 M, 36 F     \\ \hline
        \textbf{Age (yr)}                                                                                          & 27.8$\pm$19.5 & \textbf{Age range}                                                                             & 1 day-72 years \\ \hline
        \textbf{\begin{tabular}[c]{@{}l@{}}Number of subjects (age<18 years , \\ age$\geq$18 years)\end{tabular}} & 27,55         & \textit{\textbf{\begin{tabular}[c]{@{}l@{}}Number of subjects with \\ ICH\end{tabular}}}   & 36             \\ \hline
        \textbf{\begin{tabular}[c]{@{}l@{}}Number of subjects with IVH,\\  IPH, SAH, EDH, and SDH\end{tabular}}    & 5,16,7,21,4   & \textbf{\begin{tabular}[c]{@{}l@{}}Number of subjects with\\  skull fracture\end{tabular}} & 22             \\ \hline
        \end{tabular}
        \end{table}

    	\begin{figure}
    		\centering
    		\includegraphics[width=1\linewidth]{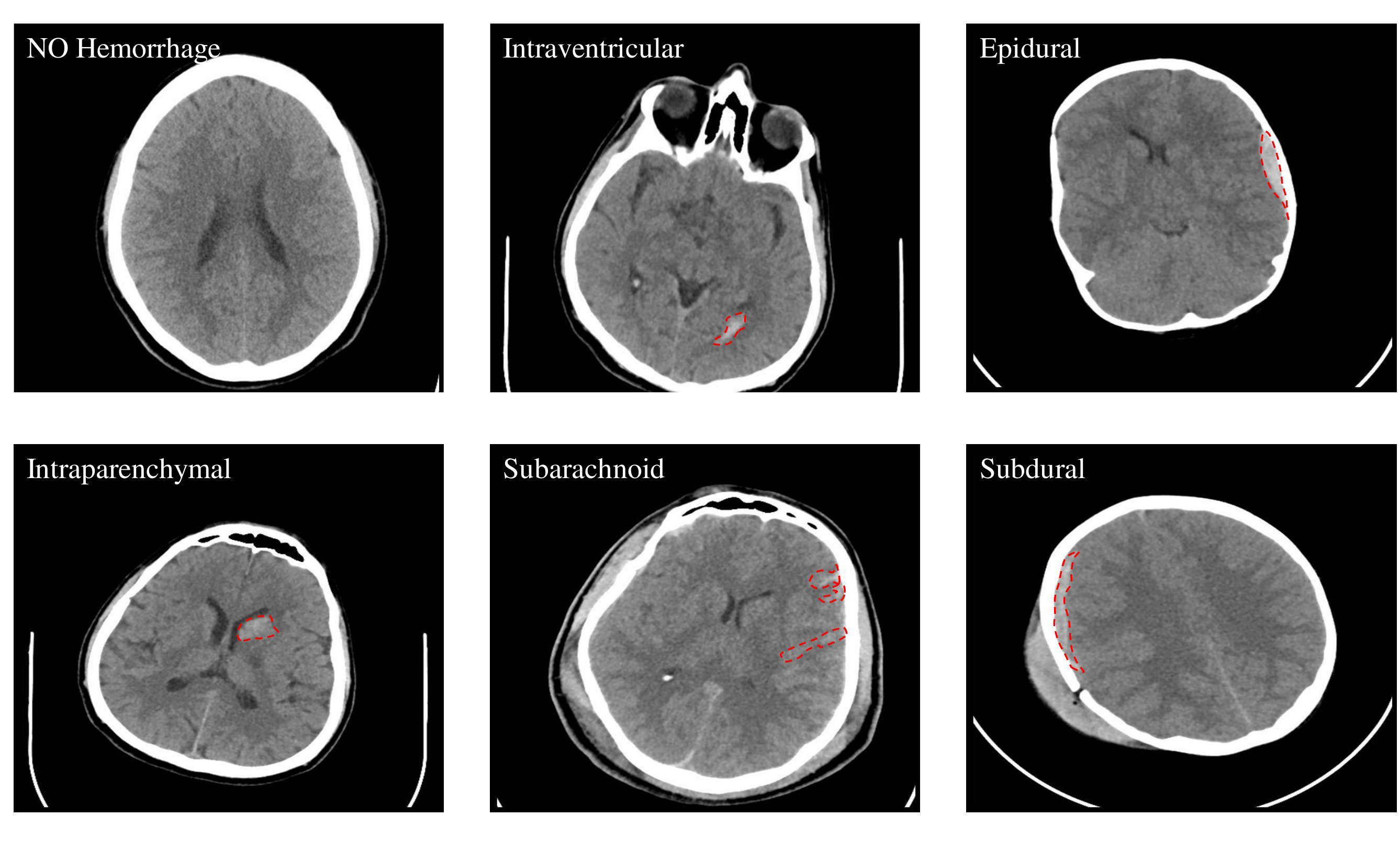}
    		\caption{Samples from the dataset that show the different types of ICH (IVH, IPH, SAH, EDH, and SDH).}
    		\label{Fig:samples}
    	\end{figure}
        
        Out of  all  the 82 subjects, 36 of the cases were diagnosed with an ICH and the following types: IVH, IPH, SAH, EDH, and SDH. See Figure \ref{Fig:samples} for some examples. One of cases had  a   chronic ICH, and it was excluded  from  this study. Table \ref{Table:NoSlices} shows the number of slices with and without an ICH as well as the numbers with different ICH sub-types.  It is important to note that  the number of  the  CT slices for each ICH sub-type in this dataset is not balanced as the majority of the CT slices do not have an ICH. Besides, IVH was only diagnosed in five subjects and  the  SDH hemorrhage in only four subjects. Also, some slices were annotated with two or more ICH sub-types. The dataset is release in JPG and \color{blue} NIfTI \color{black} formats at PhysioNet (\href{https://physionet.org/content/ct-ich/1.2.0/}{https://physionet.org/content/ct-ich/1.2.0/}), which is a repository of freely-available medical research data. The license is Creative Commons Attribution 4.0 International Public License.

    	\begin{table}[hb]
            \caption{The number of slices with and without an ICH as well as different ICH sub-types}
            \label{Table:NoSlices}
            \centering
            \begin{tabular}{|l|l|l|l|}
            %\toprule
            \hline
            \textbf{}        & \textbf{\# slices} & \textbf{}     & \textbf{\# slices} \\ \hline
            %\midrule
            Intraventricular & 24                 & Epidural      & 182                \\ \hline
            Intraparenchymal & 73                 & Subdural      & 56                 \\ \hline
            Subarachnoid     & 18                 & No Hemorrhage & 2173               \\ \hline
            %\bottomrule
            \end{tabular}
        \end{table}
    
    %%%%%%%%%%%%%%%%%%%%%%%%%%%%%%%%%%%%%%%%%%%%%%%%%%%%%%%%%%%%%%%%%%%%%%%%%%%%%%%%%%%%%%%%%%%%%%%%%%%%%%%%%%%%%%%%%%%%%%%%
    	%%%%%%%%%%%%%%%%%%%%%%%%%%%%%%%%%%%%%%%%%%%%%%%%%%%%%%%%%%%%%%%%%%%%%%%%%%%%%%%%%%%%%%%%%%%%%%%%%%%%%%%%%%%%%%%%%%%%%%%%%%%%%%%%%%%%%%%%%%%%%%%%%%%%%%%%%%%%%%%%%%%%%%%%%%%%%%%%%%%%%%%%%%%%%%%%%%%%%%%%%%%%%%%%%%%%%%%%
    	\section{ICH Segmentation Using U-Net} \label{method}
\begin{figure*} [ht]
    			\centering
    			\includegraphics[width=\linewidth]{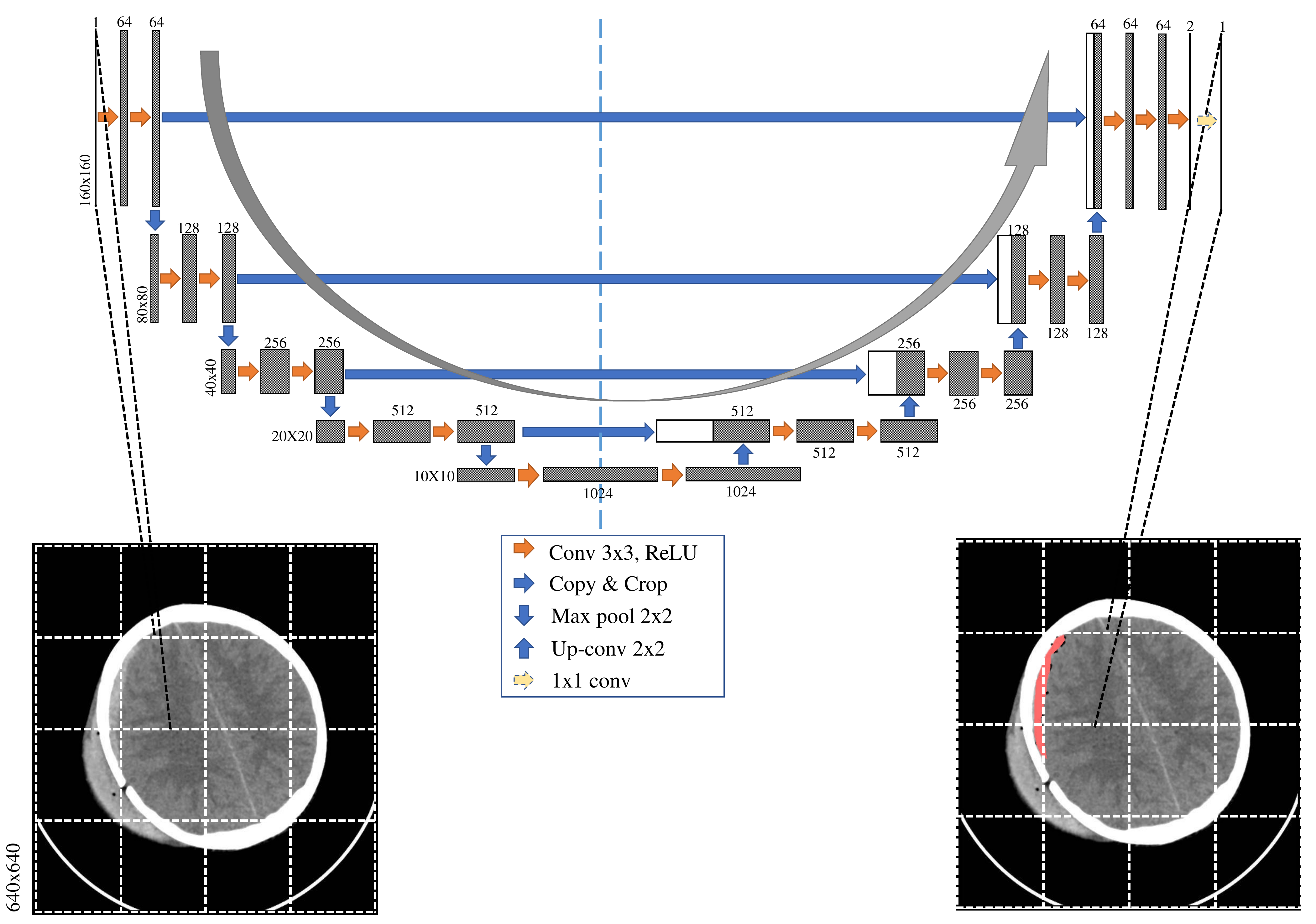}
    			\caption{The architecture of U-Net proposed in this study. Each CT slice is divided into 16 windows before feeding them to the U-Net for  the  ICH segmentation.}
    			\label{Fig:unet}
    		\end{figure*}

    	    Fully Convolutional Network (FCN) is an end-to-end or 1-stage algorithm used for semantic segmentation. Recently, FCN has exceeded the state-of-art performance in many applications involving delineation of the objects. For biomedical image segmentation, U-Net as a type of FCN was shown to be effective on small training datasets \cite{ronneberger2015u},  which motivated us to use it for the ICH segmentation in our study . In this work, we investigated the first application of U-Net for  the  ICH segmentation. The architecture of U-Net is illustrated in Figure \ref{Fig:unet}.
    	    
            The architecture is symmetrical because it builds upon two paths: a contracting path and an expansive path. In the constructing path, four blocks of typical components of a convolutional network are used. Each block is constructed by two $3\times3$ convolutional filtering layers along with padding, which is followed by a rectified linear unit (ReLU) and then by a $2\times2$ max-pooling layer. In the expansive path, four blocks are also built that consist of two $3\times3$ convolutional filtering layers followed by ReLU layers. Each block is preceded by upsampling the feature maps followed by a $2\times2$ convolution (up-convolution), which are then concatenated with the corresponding cropped feature map from the contracting path. The skip connections between the two paths are intended to provide the local or the fine-grained spatial information to the global information while upsampling for  the  precise localization. 
            After the last block in the expansive path, the feature maps are first filtered using two $3\times3$ convolutional filters to produce two images; one is for the ICH regions and one for the background. The final stage is a $1\times1$ convolutional filter with a sigmoid activation layer to produce the ICH probability in each pixel. In summary, the network has 24 convolutional layers, four max-pooling layers, four upsampling layers, and four concatenations. No dense layer is used in this architecture, in order to reduce the number of parameters and computation time.

    	%%%%%%%%%%%%%%%%%%%%%%%%%%%%%%%%%%%%%%%%%%%%%%%%%%%%%%%%%%%%%%%%%%%%%%%%%%%%%%%%%%%%%%%%%%%%%%%%%%%%%%%%%%%%%%%%%
         
    	\section{Experiments} \label{Experiments}
    	    No preprocessing was applied on the original CT slices, except removing 5 pixels from the image borders that include only the black regions. The resulted shape of the CT slices was $640\times640$. Three experiments were performed to validate the performance of U-Net and compare it with a simple threshold-based method. In the first experiment, a grid search was implemented to select the lower and upper thresholds of the ICH regions. The thresholds that resulted in the highest Jaccard index on the training data were selected and and used in the testing procedure. 
    	
    	    In the second experiment, the U-Net was trained and tested using the full $640\times640$ CT slices. However, we expected that this model will be biased to the negative class because only small number of pixels belong to the positive class in each CT scan. For the same reason, the authors in Ref. \cite{kuo_cost_sensitive_hemorrhage} used $160\times160$ crops instead of the entire CT slice and achieved a preciser model. Using this approach can also balance the training data by undersampling the negative crops. Therefore, in the third experiment, each slice from the CT scan was first divided using $160\times160$ window with an stride 80. This process resulted in 49 overlapped windows of size $160\times160$, which were then passed through U-Net for the ICH segmentation. Later, the segmented windows of each CT scan were combined to produce full $640\times640$ ICH masks. Finally, two consecutive morphological operations were performed on the ICH masks: closing to fill in the gaps in the ICH regions and opening to remove outliers and non-ICH regions.

    	     For the evaluation purposes,  we used slice-level Jaccard index (Eqn. \ref{eq1}) and Dice similarity coefficient (Eqn. \ref{eq2})  to quantify how well the model segmentation on each CT slice fits the ground truth segmentation.
            \begin{equation}
               \label{eq1} JaccardIndex=\dfrac{|{R_{ICH}\cap\hat{R_{ICH}}}|} {|{R_{ICH}\cup\hat{R_{ICH}}}|}
            \end{equation}
            \begin{equation}
               \label{eq2} Dice=\dfrac{2|{R_{ICH}\cap\hat{R_{ICH}}}|} {|R_{ICH}|+|\hat{R_{ICH}}|}
            \end{equation}
            where $R_{ICH}$ and $\hat{R_{ICH}}$ are the segmented ICH performed by the neurologists and U-Net, respectively. 
     %Note that, downsampling the CT scans may result in removing some small regions of ICH that could reduce the segmentation performance.

    	%%%%%%%%%%%%%%%%%%%%%%%%%%%%%%%%%%%%%%%%%%%%%%%%%%%%%%%%%%%%%%%%%%%%%%%%%%%%%%%%%%%%%%%%%%%%%%%%%%%%%%%%%%%%%%%%%%%%

    	\section{Results} \label{results}
            Subject-based, 5-fold cross-validation was used to train, validate, and test the developed model for all the experiments.  For the first experiment, a grid search was implemented to select a lower threshold in a 100 to 210 range, and an upper threshold in 210 to 255 range. The selected thresholds which were 140 and 230 resulted in a testing Jaccard index of 0.08 and Dice coefficient of 0.135. 
            
            For the second and third experiments, the U-Net architecture illustrated in Figure \ref{Fig:unet} was implemented in the Python environment using Keras library with TensorFlow as backend \cite{chollet2015keras}. The shape of the input image was $640\times640$ in the second experiment and $160\times160$ in the third experiment. The $640\times640$ CT slices or the $160\times160$ windows and their corresponding segmentation masks were used to train the network in each experiment.  In our dataset, 36 subjects out of 82 were diagnosed with  an  ICH, resulting in only 318 ICH slices out of 2491 (i.e., less than 10\% of the images). In order to address the class-imbalance issue, a random undersampling approach was applied to the training data to reduce the number of $640\times640$ CT slices or $160\times160$ windows that do not have  an  ICH. 
            
            At every cross-validation iteration, one fold of the CT scans was left as a held-out set for testing, one fold for validation, and three folds were used for the training purposes.  U-Net was trained for 150 epochs on the $640\times640$ CT slices or $160\times160$ windows and their corresponding segmentation windows Using GeForce RTX2080 GPU with 11 GB of memory. The training stage took approximately 5 hours in each cross-validation iteration.  During the training and at each iteration, random slices were selected from the training data, and a data augmentation was performed randomly from the following linear transformations: 
            \begin{itemize}
                \item Rotation with maximum 20 degrees
                \item Width shift with maximum 0.1\% of the image
                \item Height shift with maximum 0.1\% of the image
                \item Shear with maximum 0.1\% of the image
                \item Zoom with maximum of 0.2\% of the image
            \end{itemize}
            The dataset has a wide range of ages, which implies a wide range of head shapes and sizes, thus zooming and shearing were applied for the augmentation. Also, the head orientation could be different from subject to subject. Hence, rotation as well as width and height shifts were applied to increase the model generalizability. These linear transformations yield valid CT slices as would present in real CT data. It is worth mentioning that the non-linear deformations may provide slices that would not be seen in real CT data. As a result, we only used linear transformations in our analysis. In addition, all the subjects entered the CT scanner with their heads facing to the same direction. So the horizontal flipping will lead to CT slices that will not be generated in the data acquisition process. That is why we did not use it as an augmentation method.
    
            Adam optimizer was used with cross-entropy loss and 1e-5 learning rate.  A mini-batch of size 2 was used for the second experiment and 32 in the third experiment. The trained model was validated after each epoch. The best-trained model that resulted in the lowest validation Jaccard index was saved and used for testing purposes. The training evaluation metric was the average cross-entropy loss.
    
\begin{figure*}
            	\centering
            	\includegraphics[width=0.8\linewidth]{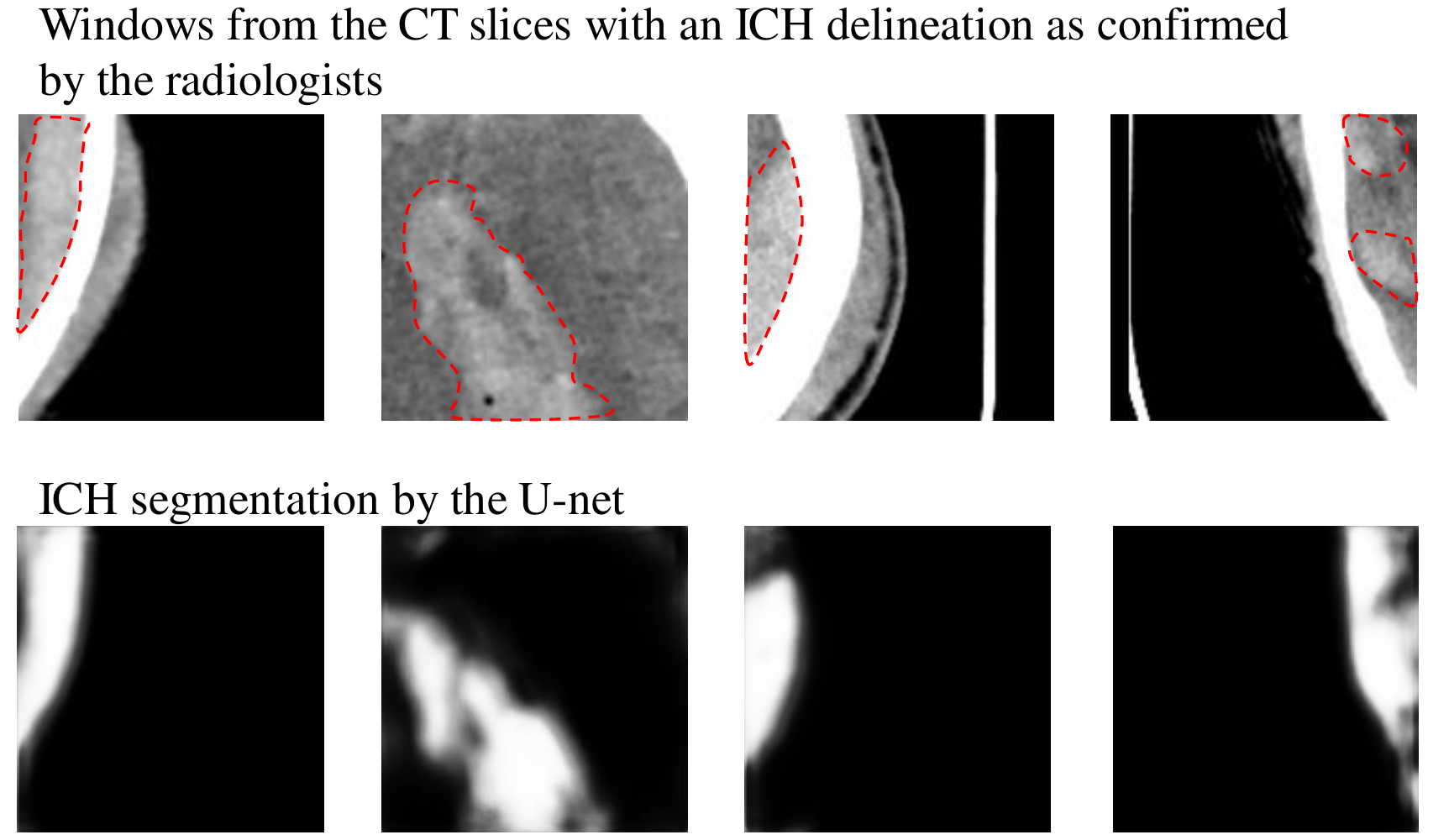}
            	\caption{Samples from the windows of the testing CT slices are shown on the top. The mask or delineation of the ICH is shown with a red dotted line. The output of U-Net before thresholding and applying the morphological operations is shown on the bottom.}
            	\label{Fig:windows_results}
            \end{figure*}

\begin{figure*}
            	\centering
            	\includegraphics[width=1\linewidth]{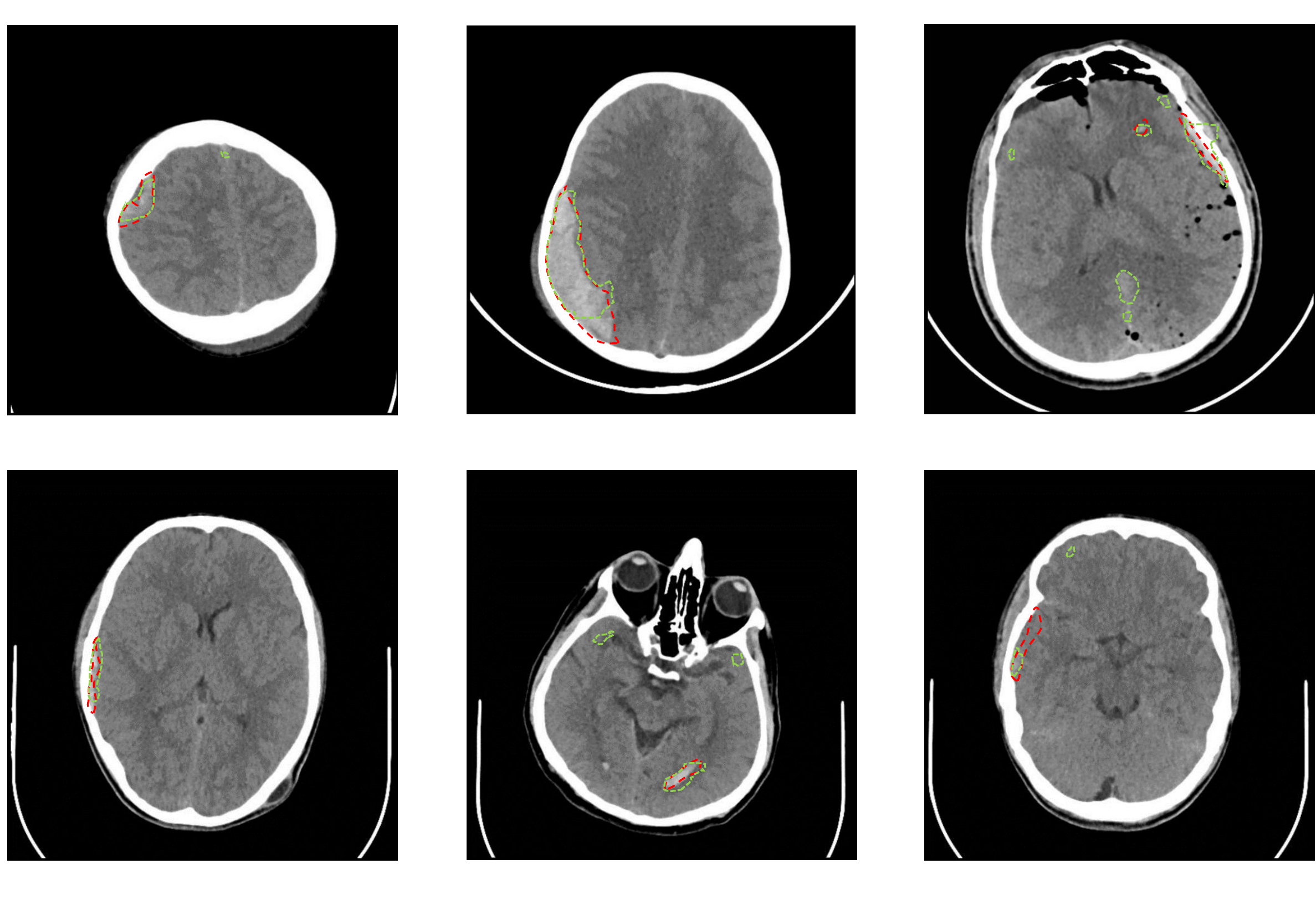}
            	\caption{Samples from the testing CT slices along with the radiologist delineation of the ICH (red dotted lines) and the U-Net segmentation (green dotted lines) are provided. A precise match of the U-Net segmentation is shown in the slices on the left side. There are some false-positive regions in the slices on the right side.}
            	\label{Fig:full_results}
            \end{figure*}

           For the second experiment when the full CT slices were used, the U-Net failed to detect any ICH region and resulted in only black masks. The reason was that even though we used only the CT slices with an ICH in the training phase, these CT slices still had very few pixels that belonged to the positive class. As a result, the training dataset was biased toward the negative class significantly. Windowing the CT slices in the third experiment improved this biasing issue by undersampling the negative crops. The 5-fold cross-validation of the developed U-Net resulted in a better performance for the third experiment as shown in Table \ref{Table:summary}. The testing Jaccard index was 0.21 and the Dice coefficient was 0.31. The slice-level, sensitivity was 97.2\% and specificity was 50.4\%. Increasing the threshold on the predicted probability masks yielded a better testing specificity at the expense of the testing sensitivity as shown in Table \ref{Table:threshold}.  Figure \ref{Fig:windows_results} provides the segmentation result of the trained U-Net on some test $160\times160$ windows along with the radiologist delineation of the ICH.  The boundary effect of each predicted $160\times160$ mask was minimal. The boundaries show low probabilities for the non-ICH regions instead of zero, and they were zeroed out after thresholding and performing the morphological operations.  The final segmented ICH regions after combining the windows, thresholding, and performing the morphological operations for some CT slices are shown in Figure \ref{Fig:full_results}. As shown in this figure, the model matched the radiologist ICH segmentation perfectly in the slices shown on the left side, but there are some false-positive ICH regions in the right-side slices.  Note that the CT slice in Figure \ref{Fig:full_results}, bottom right panel, shows the ending of an EDH region where the model only segments part of it. 
           
            \begin{table}
                \centering
                \caption{The testing results of the U-Net model trained on $160\times160$ crops and used for the ICH segmentation.}
                \label{Table:summary}
                \begin{tabular}{|l|l|l|l|l|}
                \hline
                                 & \textbf{Jaccard Index} & \textbf{Dice Coefficient} & \textbf{Sensitivity (\%)} & \textbf{Specificity (\%)} \\ \hline
                \textbf{Min}     & 0.00                   & 0.00                      & 50                        & 0                         \\ \hline
                \textbf{Max}     & 0.528                  & 0.677                     & 100                       & 100                       \\ \hline
                \textbf{STD}     & 0.163                  & 0.211                     & 9.9                       & 29.9                      \\ \hline
                \textbf{Average} & 0.218                  & 0.315                     & 97.28                     & 50.4                      \\ \hline
                \end{tabular}
            \end{table}
        
            \begin{table}
                \centering
                \caption{The testing slice-level results of the U-Net model trained on $160\times160$ crops using different thresholds.}
                \label{Table:threshold}
                \begin{tabular}{|l|l|l|l|}
                \hline
                \textbf{Threshold} & \textbf{Sensitivity (\%)} & \textbf{Specificity (\%)} & \textbf{Accuracy (\%)} \\ \hline
                \textbf{0.5}       & 97.2                 & 50.4                 & 56.6              \\ \hline
                \textbf{0.6}       & 88.7                 & 62.2                 & 65.9              \\ \hline
                \textbf{0.7}       & 77.6                 & 74.5                 & 76                \\ \hline
                \textbf{0.8}       & 73.7                 & 82.4                 & 82.5              \\ \hline
                \textbf{0.9}       & 63.1                 & 88.6                 & 87                \\ \hline
                \end{tabular}
            \end{table}
            
             The results based on the ICH sub-type showed that the U-Net performed the best with a Dice coefficient of 0.52 for the ICH segmentation of the subjects who had a SDH.  The average Dice scores for the ICH segmentation of the subjects who had an EDH, IVH, IPH and SAH were 0.35, 0.3, 0.28 and 0.23, respectively.  The minimum Dice coefficient and Jaccrad index in Table \ref{Table:summary} was zero when the U-Net failed to localize the ICH regions in the CT scans of two subjects. One of the subjects had only a small IPH region in one CT slice, and the other subject had only a small IPH region in two CT slices. The results based on the subjects' age shows that the Dice coefficient of the subjects younger than 18 years is 0.321 and for the subjects older than 18, it is 0.309. This analysis confirms that there is no significant difference between the method's performance for the subjects younger and older than 18 years. 
            %subject 79-slice 14, and subjct 58-slice 14 and 15. The IPH for these subjects where less than 10mm based on the slice thickness of 5mm.

    	\section{Discussions}
    	    A protocol was designed to collect head CT scans from subjects who had a TBI to diagnose the presence of an ICH, segment the ICH regions, and detect its sub-types. A total of 82 CT scans were collected where an ICH region was detected in 36 of them.  Later, the dataset was used to train and evaluate a threshold-based method and a U-Net network based on 5-fold cross-validation. U-Net was trained on the full CT slice in one experiment and on $160\times160$ crops in another experiment.  In the latter, each CT scan was divided into $160\times160$ overlapped windows, and an undersampling technique of the negative class (non-ICH regions) was performed to compensate for the data imbalance. 
    	    
    	     The U-Net model based on $160\times160$ crops of the CT slices resulted in a Dice coefficient of 0.31 for the ICH segmentation and a high sensitivity for detecting the ICH regions to be considered as the baseline for this dataset. This performance is comparable to the deep learning methods in the literature that were trained on small datasets \cite{nag2018computer, kuang2019segmenting}. Kuang and colleagues reported a Dice coefficient of 0.65 when a semi-automatic method based on U-Net and a contour evolution were used for the ICH segmentation. They reported a Dice coefficient of 0.35 when only U-Net was used \cite{kuang2019segmenting}. The performance of the U-Net trained in our study is comparable to their results considering that we used a smaller dataset that had all the ICH sub-types and not only intracerebral hemorrhage. Also, \cite{nag2018computer} tested autoencoder and active contour Chan-Vese model on a dataset that did not contain any SDH cases and reported an average Jaccard index of 0.55. The autoencoder was trained on half of the dataset, and later all the dataset was used for testing, which could boost the average Jaccard index.  The other deep learning-based models in Ref. \cite{chang2018hybrid, lee2019explainable, kuo_cost_sensitive_hemorrhage, cho2019improving} were trained and tested on larger datasets and achieved higher performance for the ICH segmentation. Ref. \cite{chang2018hybrid} reported an average Dice coefficient of 0.85, Ref. \cite{lee2019explainable} reported a 78\% overlap between the attention maps of their CNN model and the gold-standard bleeding points, Ref.  \cite{kuo_cost_sensitive_hemorrhage} reported 78\% average precision, and \cite{cho2019improving} reported 80.19\% precision and 82.15\% recall.  In addition to the deep learning methods, in the study of Ref. \cite{shahangian2016automatic}, DRLSE was used for the segmentation of EDH, IPH, and SDH, and Dice coefficients of 0.75, 0.62 and 0.37 were reported for each sub-type, respectively. Our method achieved a higher Dice coefficient of 0.52 in segmenting SDH. Some traditional methods reported better dice coefficient (0.87 \cite{bhadauria2014intracranial}, 0.89 \cite{muschelli2017pitchperfect}, and 0.82 \cite{ gautam2019automatic}) for the ICH segmentation when a small dataset was used. %The software of the method implemented by Muschelli and colleagues is available online, but it requires the raw CT scan data, while our data consists of the JPG format of the CT scans with no access to the DICOM files. Hence, we cannot apply their algorithm on our dataset.

    	     Regarding the ICH detection, U-Net achieved a slice-level sensitivity of 97.2\% and specificity of 50.4\%, which is comparable to the results reported by Yuh and colleagues \cite{yuh2008computer} when 0.5 threshold was used. Increasing the threshold to 0.8 resulted in 73.7\% sensitivity, 82.4\% specificity, and 82.5\% accuracy, which is comparable to some methods in the literature that were trained on large datasets \cite{arbabshirani2018advanced, grewal2018radnet}. In \cite{arbabshirani2018advanced}, an ensemble of four 3D CNN models was trained on 10k CT scans and yielded 71.5\% sensitivity and 83.5\% specificity. In \cite{grewal2018radnet}, a deep model based on DenseNet and RNN achieved 81\% accuracy. 
    	    
    	     Our observation was that the main reason for the low Dice coefficient of the trained U-Net was the false positive segmentation as shown in Figure \ref{Fig:full_results}. The false positive segmentation was more prevalent near the bones where the intensity in the grayscale image is similar to the intensity of the ICH region. Another limitation is that the developed U-Net model failed to localize the ICH regions in the CT scans of two subjects who had a small IPH region. Hence, the current method as stands can be used as an assistive technology to the radiologists for the ICH segmentation but is not yet at a precision that can be used as a standalone segmentation method.  Future work could be collecting further data and also  enhancing U-Net with a recurrent neural network such as LSTM networks to consider the relationship between the adjacent scans when segmenting the ICH regions. Besides, the performance can be improved by utilizing a transfer learning to initialize the model weights before training the model on the ICH small dataset.

    	\section{Conclusions}
        ICH is a critical medical lesion that requires an immediate medical attention, or it may turn into a secondary brain insult that could lead to paralysis or even death. The contribution of this paper is two-fold. First, a new dataset with 82 CT scans was collected. The dataset is made publicly available online at Physionet to address the need for more publicly available benchmark datasets toward developing reliable techniques for the automated ICH segmentation. Second, a deep learning method for the ICH segmentation was developed. The developed method was assessed on the collected data with 5-fold cross-validation. It resulted in  a Dice coefficient of 0.31, which has a comparative performance for deep learning methods reported in the literature  using  small datasets.  Moreover, the paper provides a detailed review of the methods for  the  detection of ICH and its sub-types as well as segmentation of  the  ICH. Developing an automated ICH screening tool could improve the diagnosis and management of ICH significantly when experts are not immediately available in the emergency rooms, especially in developing countries or remote areas.

%%%%%%%%%%%%%%%%%%%%%%%%%%%%%%%%%%%%%%%%%%
%% optional
%\supplementary{The following are available online at \linksupplementary{s1}, Figure S1: title, Table S1: title, Video S1: title.}


%%%%%%%%%%%%%%%%%%%%%%%%%%%%%%%%%%%%%%%%%%
\authorcontributions{Conceptualization, Murtadha Hssayeni, Muayad Croock, Aymen Al-Ani, Hassan Al-khafaji, and Behnaz Ghoraani; Data curation, Murtadha Hssayeni, and Zakaria Yahya; Formal analysis, Murtadha Hssayeni; Investigation, Hassan Al-khafaji, and Behnaz Ghoraani; Methodology, Murtadha Hssayeni, Muayad Croock, Aymen Al-Ani, Hassan Al-khafaji, and Behnaz Ghoraani; Resources, Hassan Al-khafaji, Zakaria Yahya, and Behnaz Ghoraani; Software, Murtadha Hssayeni; Validation, Murtadha Hssayeni; Writing – original draft, Murtadha Hssayeni and Behnaz Ghoraani; Writing – review and editing, Murtadha Hssayeni, Muayad Croock, Aymen Al-Ani and Behnaz Ghoraani.}

%\authorcontributions{conceptualization, M.D.H., M.S.C., A.A. and H.F.A.; methodology, M.D.H., M.S.C., A.A. and H.F.A.; software, M.D.H.; validation, M.D.H. and B.G.; formal analysis, M.D.H.; investigation, M.D.H. and B.G. and H.F.A; Data Annotation, H.F.A. and Z.A.Y.; data curation, M.D.H., H.F.A. and Z.A.Y.; writing--original draft preparation, M.D.H. and B.G.; writing--review and editing, M.D.H., B.G., M.S.C., A.A. and H.F.A.; visualization, X.X.; supervision, X.X.; project administration, X.X.; funding acquisition, Y.Y.'', please turn to the  \href{http://img.mdpi.org/data/contributor-role-instruction.pdf}{CRediT taxonomy} for the term explanation. Authorship must be limited to those who have contributed substantially to the work reported.}

%%%%%%%%%%%%%%%%%%%%%%%%%%%%%%%%%%%%%%%%%%
\funding{This research received no external funding.}

%%%%%%%%%%%%%%%%%%%%%%%%%%%%%%%%%%%%%%%%%%
\acknowledgments{Thanks for Mohammed Ali for the clinical support and all the subjects participated in the data collection.}

%%%%%%%%%%%%%%%%%%%%%%%%%%%%%%%%%%%%%%%%%%
\conflictsofinterest{The authors declare no conflict of interest.} 

%%%%%%%%%%%%%%%%%%%%%%%%%%%%%%%%%%%%%%%%%%
%% optional
\abbreviations{The following abbreviations are used in this manuscript:\\

\noindent 
\begin{tabular}{@{}ll}
CT & Computerized Tomography\\
TBI & Traumatic brain injury\\
ICH & Intracranial hemorrhage\\
IVH & Intraventricular hemorrhage\\
IPH & Intraparenchymal hemorrhage\\
SAH & Subarachnoid hemorrhage\\
EDH & Epidural hemorrhage\\
SDH & Subdural hemorrhage\\
CNN & Convolutional neural networks\\
RNN & Recurrent neural network\\
FCN & Fully convolutional networks\\
LSTM & Long short-term memory network\\
AUC & Area under the ROC curve\\
\end{tabular}}

%%%%%%%%%%%%%%%%%%%%%%%%%%%%%%%%%%%%%%%%%%
%% optional
%\appendixtitles{no} %Leave argument "no" if all appendix headings stay EMPTY (then no dot is printed after "Appendix A"). If the appendix sections contain a heading then change the argument to "yes".
%\appendix
%\section{}
%\unskip
%\subsection{}
%The appendix is an optional section that can contain details and data supplemental to the main text. For example, explanations of experimental details that would disrupt the flow of the main text, but nonetheless remain crucial to understanding and reproducing the research shown; figures of replicates for experiments of which representative data is shown in the main text can be added here if brief, or as Supplementary data. Mathematical proofs of results not central to the paper can be added as an appendix.

%\section{}
%All appendix sections must be cited in the main text. In the appendixes, Figures, Tables, etc. should be labeled starting with `A', e.g., Figure A1, Figure A2, etc. 

%%%%%%%%%%%%%%%%%%%%%%%%%%%%%%%%%%%%%%%%%%
%% optional
%\sampleavailability{Samples of the compounds ...... are available from the authors.}

%% for journal Sci
%\reviewreports{\\
%Reviewer 1 comments and authors’ response\\
%Reviewer 2 comments and authors’ response\\
%Reviewer 3 comments and authors’ response
%}

%%%%%%%%%%%%%%%%%%%%%%%%%%%%%%%%%%%%%%%%%%

\reftitle{References}
\bibliography{Refs}

\begin{thebibliography}{-------}
\providecommand{\natexlab}[1]{#1}

\bibitem[Taylor \em{et~al.}(2017)Taylor, Bell, Breiding, and
  Xu]{taylor2017traumatic}
Taylor, C.A.; Bell, J.M.; Breiding, M.J.; Xu, L.
\newblock Traumatic Brain Injury-Related Emergency Department Visits,
  Hospitalizations, and Deaths-United States, 2007 and 2013.
\newblock {\em Morbidity and mortality weekly report. Surveillance summaries
  (Washington, DC: 2002)} {\bf 2017}, {\em 66},~1--16.

\bibitem[van Asch \em{et~al.}(2010)van Asch, Luitse, Rinkel, van~der Tweel,
  Algra, and Klijn]{van2010incidence}
van Asch, C.J.; Luitse, M.J.; Rinkel, G.J.; van~der Tweel, I.; Algra, A.;
  Klijn, C.J.
\newblock Incidence, case fatality, and functional outcome of intracerebral
  haemorrhage over time, according to age, sex, and ethnic origin: a systematic
  review and meta-analysis.
\newblock {\em The Lancet Neurology} {\bf 2010}, {\em 9},~167--176.

\bibitem[Currie \em{et~al.}(2016)Currie, Saleem, Straiton, Macmullen-Price,
  Warren, and Craven]{currie2016imaging}
Currie, S.; Saleem, N.; Straiton, J.A.; Macmullen-Price, J.; Warren, D.J.;
  Craven, I.J.
\newblock Imaging assessment of traumatic brain injury.
\newblock {\em Postgraduate medical journal} {\bf 2016}, {\em 92},~41--50.

\bibitem[Xue \em{et~al.}(2012)Xue, Antani, Long, Demner-Fushman, and
  Thoma]{xue2012window}
Xue, Z.; Antani, S.; Long, L.R.; Demner-Fushman, D.; Thoma, G.R.
\newblock Window classification of brain CT images in biomedical articles.
\newblock  AMIA Annual Symposium Proceedings. American Medical Informatics
  Association,  2012, Vol. 2012, p. 1023.

\bibitem[Litjens \em{et~al.}(2017)Litjens, Kooi, Bejnordi, Setio, Ciompi,
  Ghafoorian, Van Der~Laak, Van~Ginneken, and S{\'a}nchez]{litjens2017survey}
Litjens, G.; Kooi, T.; Bejnordi, B.E.; Setio, A.A.A.; Ciompi, F.; Ghafoorian,
  M.; Van Der~Laak, J.A.; Van~Ginneken, B.; S{\'a}nchez, C.I.
\newblock A survey on deep learning in medical image analysis.
\newblock {\em Medical image analysis} {\bf 2017}, {\em 42},~60--88.

\bibitem[Chilamkurthy \em{et~al.}(2018)Chilamkurthy, Ghosh, Tanamala, Biviji,
  Campeau, Venugopal, Mahajan, Rao, and Warier]{chilamkurthy2018deep}
Chilamkurthy, S.; Ghosh, R.; Tanamala, S.; Biviji, M.; Campeau, N.G.;
  Venugopal, V.K.; Mahajan, V.; Rao, P.; Warier, P.
\newblock Deep learning algorithms for detection of critical findings in head
  CT scans: a retrospective study.
\newblock {\em The Lancet} {\bf 2018}.

\bibitem[Chan(2007)]{chan2007computer}
Chan, T.
\newblock Computer aided detection of small acute intracranial hemorrhage on
  computer tomography of brain.
\newblock {\em Computerized Medical Imaging and Graphics} {\bf 2007}, {\em
  31},~285--298.

\bibitem[Yuh \em{et~al.}(2008)Yuh, Gean, Manley, Callen, and
  Wintermark]{yuh2008computer}
Yuh, E.L.; Gean, A.D.; Manley, G.T.; Callen, A.L.; Wintermark, M.
\newblock Computer-aided assessment of head computed tomography (CT) studies in
  patients with suspected traumatic brain injury.
\newblock {\em Journal of neurotrauma} {\bf 2008}, {\em 25},~1163--1172.

\bibitem[Li \em{et~al.}(2012{\natexlab{a}})Li, Wu, Li, Li, Du, Chen, Jia, and
  Hu]{li2010automatic}
Li, Y.; Wu, J.; Li, H.; Li, D.; Du, X.; Chen, Z.; Jia, F.; Hu, Q.
\newblock Automatic detection of the existence of subarachnoid hemorrhage from
  clinical CT images.
\newblock {\em Journal of medical systems} {\bf 2012}, {\em 36},~1259--1270.

\bibitem[Li \em{et~al.}(2012{\natexlab{b}})Li, Zhang, Hu, Li, Jia, and
  Wu]{li2012automatic}
Li, Y.H.; Zhang, L.; Hu, Q.M.; Li, H.W.; Jia, F.C.; Wu, J.H.
\newblock Automatic subarachnoid space segmentation and hemorrhage detection in
  clinical head CT scans.
\newblock {\em International journal of computer assisted radiology and
  surgery} {\bf 2012}, {\em 7},~507--516.

\bibitem[Shahangian and Pourghassem(2016)]{shahangian2016automatic}
Shahangian, B.; Pourghassem, H.
\newblock Automatic brain hemorrhage segmentation and classification algorithm
  based on weighted grayscale histogram feature in a hierarchical
  classification structure.
\newblock {\em Biocybernetics and Biomedical Engineering} {\bf 2016}, {\em
  36},~217--232.

\bibitem[Prevedello \em{et~al.}(2017)Prevedello, Erdal, Ryu, Little, Demirer,
  Qian, and White]{prevedello2017automated}
Prevedello, L.M.; Erdal, B.S.; Ryu, J.L.; Little, K.J.; Demirer, M.; Qian, S.;
  White, R.D.
\newblock Automated critical test findings identification and online
  notification system using artificial intelligence in imaging.
\newblock {\em Radiology} {\bf 2017}, {\em 285},~923--931.

\bibitem[Grewal \em{et~al.}(2018)Grewal, Srivastava, Kumar, and
  Varadarajan]{grewal2018radnet}
Grewal, M.; Srivastava, M.M.; Kumar, P.; Varadarajan, S.
\newblock RADnet: Radiologist level accuracy using deep learning for hemorrhage
  detection in CT scans.
\newblock  Biomedical Imaging (ISBI 2018), 2018 IEEE 15th International
  Symposium on. IEEE,  2018, pp. 281--284.

\bibitem[Jnawali \em{et~al.}(2018)Jnawali, Arbabshirani, Rao, and
  Patel]{jnawali2018deep}
Jnawali, K.; Arbabshirani, M.R.; Rao, N.; Patel, A.A.
\newblock Deep 3D convolution neural network for CT brain hemorrhage
  classification.
\newblock  Medical Imaging 2018: Computer-Aided Diagnosis. International
  Society for Optics and Photonics,  2018, Vol. 10575, p. 105751C.

\bibitem[Chang \em{et~al.}(2018)Chang, Kuoy, Grinband, Weinberg, Thompson,
  Homo, Chen, Abcede, Shafie, Sugrue, et~al.]{chang2018hybrid}
Chang, P.; Kuoy, E.; Grinband, J.; Weinberg, B.; Thompson, M.; Homo, R.; Chen,
  J.; Abcede, H.; Shafie, M.; Sugrue, L.; others.
\newblock Hybrid 3D/2D convolutional neural network for hemorrhage evaluation
  on head CT.
\newblock {\em American Journal of Neuroradiology} {\bf 2018}, {\em
  39},~1609--1616.

\bibitem[Arbabshirani \em{et~al.}(2018)Arbabshirani, Fornwalt, Mongelluzzo,
  Suever, Geise, Patel, and Moore]{arbabshirani2018advanced}
Arbabshirani, M.R.; Fornwalt, B.K.; Mongelluzzo, G.J.; Suever, J.D.; Geise,
  B.D.; Patel, A.A.; Moore, G.J.
\newblock Advanced machine learning in action: identification of intracranial
  hemorrhage on computed tomography scans of the head with clinical workflow
  integration.
\newblock {\em npj Digital Medicine} {\bf 2018}, {\em 1},~9.

\bibitem[Lee \em{et~al.}(2019)Lee, Yune, Mansouri, Kim, Tajmir, Guerrier,
  Ebert, Pomerantz, Romero, Kamalian, et~al.]{lee2019explainable}
Lee, H.; Yune, S.; Mansouri, M.; Kim, M.; Tajmir, S.H.; Guerrier, C.E.; Ebert,
  S.A.; Pomerantz, S.R.; Romero, J.M.; Kamalian, S.; others.
\newblock An explainable deep-learning algorithm for the detection of acute
  intracranial haemorrhage from small datasets.
\newblock {\em Nature Biomedical Engineering} {\bf 2019}, {\em 3},~173.

\bibitem[Ye \em{et~al.}(2019)Ye, Gao, Yin, Guo, Zhao, Lu, Wang, Bai, Cao, Song,
  et~al.]{ye2019precise}
Ye, H.; Gao, F.; Yin, Y.; Guo, D.; Zhao, P.; Lu, Y.; Wang, X.; Bai, J.; Cao,
  K.; Song, Q.; others.
\newblock Precise diagnosis of intracranial hemorrhage and subtypes using a
  three-dimensional joint convolutional and recurrent neural network.
\newblock {\em European radiology} {\bf 2019}, pp. 1--11.

\bibitem[Cho \em{et~al.}(2019)Cho, Park, Karki, Lee, Ko, Kim, Lee, Choe, Son,
  Kim, et~al.]{cho2019improving}
Cho, J.; Park, K.S.; Karki, M.; Lee, E.; Ko, S.; Kim, J.K.; Lee, D.; Choe, J.;
  Son, J.; Kim, M.; others.
\newblock Improving Sensitivity on Identification and Delineation of
  Intracranial Hemorrhage Lesion Using Cascaded Deep Learning Models.
\newblock {\em Journal of digital imaging} {\bf 2019}, {\em 32},~450--461.

\bibitem[Prakash \em{et~al.}(2012)Prakash, Zhou, Morgan, Hanley, and
  Nowinski]{prakash2012segmentation}
Prakash, K.B.; Zhou, S.; Morgan, T.C.; Hanley, D.F.; Nowinski, W.L.
\newblock Segmentation and quantification of intra-ventricular/cerebral
  hemorrhage in CT scans by modified distance regularized level set evolution
  technique.
\newblock {\em International journal of computer assisted radiology and
  surgery} {\bf 2012}, {\em 7},~785--798.

\bibitem[Bhadauria and Dewal(2014)]{bhadauria2014intracranial}
Bhadauria, H.; Dewal, M.
\newblock Intracranial hemorrhage detection using spatial fuzzy c-mean and
  region-based active contour on brain CT imaging.
\newblock {\em Signal, Image and Video Processing} {\bf 2014}, {\em
  8},~357--364.

\bibitem[Muschelli \em{et~al.}(2017)Muschelli, Sweeney, Ullman, Vespa, Hanley,
  and Crainiceanu]{muschelli2017pitchperfect}
Muschelli, J.; Sweeney, E.M.; Ullman, N.L.; Vespa, P.; Hanley, D.F.;
  Crainiceanu, C.M.
\newblock PItcHPERFeCT: Primary intracranial hemorrhage probability estimation
  using random forests on CT.
\newblock {\em NeuroImage: Clinical} {\bf 2017}, {\em 14},~379--390.

\bibitem[Kuo \em{et~al.}(2018)Kuo, H{\"a}ne, Yuh, Mukherjee, and
  Malik]{kuo_cost_sensitive_hemorrhage}
Kuo, W.; H{\"a}ne, C.; Yuh, E.; Mukherjee, P.; Malik, J.
\newblock Cost-Sensitive Active Learning for Intracranial Hemorrhage Detection.
\newblock  Medical Image Computing and Computer Assisted Intervention -- MICCAI
  2018; Frangi, A.F.; Schnabel, J.A.; Davatzikos, C.; Alberola-L{\'o}pez, C.;
  Fichtinger, G., Eds.; Springer International Publishing: Cham,  2018; pp.
  715--723.

\bibitem[Nag \em{et~al.}(2018)Nag, Chatterjee, Sadhu, Chatterjee, and
  Ghosh]{nag2018computer}
Nag, M.K.; Chatterjee, S.; Sadhu, A.K.; Chatterjee, J.; Ghosh, N.
\newblock Computer-assisted delineation of hematoma from CT volume using
  autoencoder and Chan Vese model.
\newblock {\em International journal of computer assisted radiology and
  surgery} {\bf 2018}, pp. 1--11.

\bibitem[Kuang \em{et~al.}(2019)Kuang, Menon, and Qiu]{kuang2019segmenting}
Kuang, H.; Menon, B.K.; Qiu, W.
\newblock Segmenting Hemorrhagic and Ischemic Infarct Simultaneously From
  Follow-Up Non-Contrast CT Images in Patients With Acute Ischemic Stroke.
\newblock {\em IEEE Access} {\bf 2019}, {\em 7},~39842--39851.

\bibitem[Gautam and Raman(2019)]{gautam2019automatic}
Gautam, A.; Raman, B.
\newblock Automatic Segmentation of Intracerebral Hemorrhage from Brain CT
  Images. In {\em Machine Intelligence and Signal Analysis}; Springer,  2019;
  pp. 753--764.

\bibitem[Chi \em{et~al.}(2014)Chi, Lang, Sun, Tang, Xu, Zheng, and
  Zhao]{chi2014relationship}
Chi, F.l.; Lang, T.c.; Sun, S.j.; Tang, X.j.; Xu, S.y.; Zheng, H.b.; Zhao, H.s.
\newblock Relationship between different surgical methods, hemorrhage position,
  hemorrhage volume, surgical timing, and treatment outcome of hypertensive
  intracerebral hemorrhage.
\newblock {\em World journal of emergency medicine} {\bf 2014}, {\em 5},~203.

\bibitem[Strub \em{et~al.}(2007)Strub, Leach, Tomsick, and
  Vagal]{strub2007overnight}
Strub, W.; Leach, J.; Tomsick, T.; Vagal, A.
\newblock Overnight preliminary head CT interpretations provided by residents:
  locations of misidentified intracranial hemorrhage.
\newblock {\em American Journal of Neuroradiology} {\bf 2007}, {\em
  28},~1679--1682.

\bibitem[Ronneberger \em{et~al.}(2015)Ronneberger, Fischer, and
  Brox]{ronneberger2015u}
Ronneberger, O.; Fischer, P.; Brox, T.
\newblock U-net: Convolutional networks for biomedical image segmentation.
\newblock  International Conference on Medical image computing and
  computer-assisted intervention. Springer,  2015, pp. 234--241.

\bibitem[Chollet \em{et~al.}(2015)Chollet et~al.]{chollet2015keras}
Chollet, F.; others.
\newblock Keras.
\newblock \url{https://keras.io},  2015.

\end{thebibliography}
    	
\end{document}